\documentclass[prl,twocolumn,showpacs,floatfix,amsfonts]{revtex4}
\usepackage{graphicx,graphics,color,epsfig}
\usepackage{bm}
\usepackage{amsmath}
\usepackage{amssymb}
\begin{document}
\preprint{}
\title{Dopant-Induced Local Pairing Inhomogeneity in
 Bi$_2$Sr$_2$CaCu$_2$O$_{8+\delta}$}
\author{Jian-Xin Zhu}

\affiliation{Theoretical Division, Los Alamos National
Laboratory, Los Alamos, New Mexico 87545, USA}

\date{September 6, 2005}

\begin{abstract}
A new theoretical model is presented to study the nanoscale
electronic inhomogeneity in high-$T_c$ cuprates. In this model, we
argue that the randomly distributed out-of-plane interstitial oxygen
dopants induces locally the off-diagonal (i.e., hopping integral)
disorder. This disorder modulates the superexchange interaction
resulting from a large-$U$ Hubbard model, which in turns changes the
local pairing interaction. The microscopic self-consistent
calculations shows that the large gap regions are registered to the
locations of dopants. Large gap regions exhibit small and broader
coherence peaks. These results are qualitatively consistent with
recent STM observations on optimally doped
Bi$_2$Sr$_2$CaCu$_2$O$_{8+\delta}$.

\end{abstract}


\maketitle

One of the most striking features of high-$T_c$ cuprate,
Bi$_{2}$Sr$_{2}$CaCu$_{2}$O$_{8+\delta}$ (BSCCO), is the nanoscale
electronic inhomogeneity as observed, for example, by the scanning
tunneling microscopy (STM)~\cite{Cren00,
Howard01,Pan01,Lang02,Fang04,McElroy05}. Earlier STM experiments
shows that the spectra gap varies dramatically on a very short
length scale, typically in the scale  of ~50$\AA$. At the same time
the low energy part of the tunneling  spectra are extremely
homogeneous spatially, except in the immediate vicinity of known
defects in the CuO$_{2}$
plane~\cite{SHPan00,EWHudson01,AVBalatsky04}. Coherence peaks in
regions with large gap are broadened and their height is reduced or
peaks are entirely absent. Several scenarios have been proposed to
understanding this surprising electronic inhomogeneity. It is
commonly accepted  that the hole doping in the CuO$_2$ planes  is
proportional to the concentration of out-of-plane oxygen dopant
atoms. Thus the  out-of-plane doping  also introduces disorder.
Therefore, it is tempting to speculate that poorly screened
electrostratic potentials of the dopant atoms cause a variation in
the local doping concentration, giving rise to the gap
modulations~\cite{IMartin01,ZWang02,QHWang02}. Alternatively, one
would ascribe the distinct electronic structures to the competition
between two different electronic orders that exist intrisically in
the system--- In the small gap regions, with high intensity sharp
coherence  peaks, the superconductivity wins; in the large gap
regions, a pseudo-gap phase dominates with a competing order such as
spin~\cite{SAKivelson03,WAAtkinson05,GAlvarez05} or
orbital~\cite{AGhosal04} antiferromagnetism, or charge density
wave~\cite{DPodolsky03}. The competing order scenario is supported
by the STM data on the underdoped
samples~\cite{MVershinin04,KMcElroy05} or on the optimally doped
ones in the mixed state~\cite{JEHoffman02a}, where a checkerboard
pattern in the local density of states (LDOS) has been observed. One
distinctive feature of all these scenarios is that  a uniform
pairing interaction throughout the system has always been assumed.

Recent STM data on the optimally doped BSCCO~\cite{McElroy05} showed
that the locations of oxygen dopants are positively correlated with
the large gap regions with small charge density variations.   It was
proposed by Nunner {\em et al.}~\cite{Nunner05} that the dopant
atoms modulate the local pairing interaction, i.e., the local
attractive coupling $g$ between electrons are spatially dependent.
In the phenomenological model of~\cite{Nunner05}, the modulation of
$g$ is suggested to arise from a modification of local electronic
structure parameters, but no concrete microscopic model was
explored.

In this Letter we propose a microscopic model for the pair potential
disorder. We argue that it originates from the modulation of the
local pairing interaction induced by the doped oxygen atoms. The
specific mechanism is  based on an extended $t$-$J$ model with the
local superexchange interaction being modulated by the off-diagonal
disorder (i.e., the electronic hopping integral). This specific
mechanism on how  the gap inhomogeneity might be triggered by the
disorder in pair potential, is also consistent with the
findings~\cite{JXZhu05,ACFang05} that $\tau_1$-type impurities
rather than $\tau_3$-type impurities in the Nambu space are more
relevant in the interpretation of the $\mathbf{q}$ structure
observed in most Fourier-transformed STM
experiments~\cite{JEHoffman02b,KMcElroy03,JLee05}.

We find that the off-diagonal disorder, triggered by dopant atoms,
produces a modulations of the gap across the sample. To be
consistent with the experiment, we will assume
 that dopant atoms are interstitial.  Each dopant
atom therefore leads to  a local modulation of the hopping matrix
element on the neighboring links in Cu-O plane. This increased
hopping will elevate the local superexchange interaction, which in
turn leads to the local increase of the pair potential. Typical
length scale of the modulation in response to the electronic
distortion is set by the superconducting coherence length $\xi \sim
5a$ and is short in our model. We also find that the gap exhibited
in the LDOS spectra reflects directly the local pair potential; and
that the LDOS spectra in the regions of the large gap have coherence
peaks that are moving out and are less sharp than the coherence
peaks at small-gap regions. We believe the effect found here is a
generic property of any model, where the pairing scale is set by the
local superexchange.

As early as in 1987, Anderson suggested the relevance of the
large-$U$ limit of the Hubbard model to the problem of high-$T_c$
superconductivity in cuprates~\cite{PWAnderson87} and the strong
correlations are a central part of the high-$T_c$ problem. To attack
the present problem, we therefore start with a large-$U$ Hubbard
model with an off-diagonal disorder, defined on a two-dimensional
square lattice:
\begin{equation}
\mathcal{H} = -\sum_{\langle \mathbf{ij}\rangle, \sigma} t_{\bf ij}
c_{\mathbf{i} \sigma}^{\dagger}c_{\mathbf{j}\sigma} +
\sum_{\mathbf{i}} (\epsilon_{\mathbf{i}}-\mu)
c_{\mathbf{i}\sigma}^{\dagger} c_{\mathbf{i}\sigma}
+U\sum_{\mathbf{i}} n_{\mathbf{i}\uparrow}n_{\mathbf{i}\downarrow}
\;.
\end{equation}
Here the quantities $c_{\mathbf{i}\sigma}^{(\dagger)}$ annihilates
(creates) an electron on the $\mathbf{i}$th site with spin $\sigma$,
$t_{\mathbf{ij}}$ is the hopping integral between between sites
$\mathbf{i}$ and $\mathbf{j}$, $\epsilon_{\mathbf{i}}$ is the
on-site energy, $\mu$ is the chemical potential,
$n_{\mathbf{i}\sigma}$ is the spin-polarized number operator, and
$\sum_{\langle \mathbf{ij}\rangle}$ indicates the summation over
neighboring sites. Following the similar treatment to a large-$U$
Hubbard model for a clean system~\cite{JEHirsh85,CGros87,FCZhang88},
one can show that, in the large-$U$ limit, the disordered Hubbard
model is equivalent to the following model Hamiltonian:
\begin{eqnarray}
{\cal H} &=& -\sum_{\langle \mathbf{ij}\rangle, \sigma}
t_{\bf ij} c_{\mathbf{i}
\sigma}^{\dagger}c_{\mathbf{j}\sigma} + \sum_{\mathbf{i}}
(\epsilon_{\mathbf{i}}-\mu) c_{\mathbf{i}\sigma}^{\dagger}
c_{\mathbf{i}\sigma} \nonumber \\
&& +\frac{1}{2} \sum_{\langle \mathbf{ij}\rangle} J_{\mathbf{ij}}[
\mathbf{S}_{\mathbf{i}}\cdot \mathbf{S}_{\mathbf{j}}-\frac{1}{4}
n_{\bf i}n_{\bf j}] \;, \label{EQ:t-J}
\end{eqnarray}
acting on the subspace of empty and singly-occupied sites only, where
 $\mathbf{S}_{\mathbf{i}}$ is
the Heisenberg spin-$\frac{1}{2}$ operator, the local superexchange
interaction $J_{\mathbf{ij}}=4t_{\mathbf{ij}}^{2}/U$.  In
traditional disordered Hubbard model, the disorder is modeled
through a random distribution of on-site (diagonal) energy while the
hopping integrals are taken to be uniform. This approach cannot
generate a direct modulation of the local pairing interaction, and
the resultant pair potential can only be modulated near the known
impurities through a local shift in effective chemical potential
within a self-consistent solution.

Here we argue instead that the oxygen dopants modulate the local
electronic hopping integral, and as a direct consequence, modulate
the local pairing interaction through  superexchange
$J_{\mathbf{ij}}$. For the  purpose of this discussion, we consider
only the optimally doped regime. In this regime (and hopefully also
in the overdoped regime), we believe that the
 slave-boson mean field approach might be a reasonable
 approximation~\cite{GKotliar88}. In this approach, one first
writes the electron operator as $c_{\mathbf{i}\sigma} =
b_{\mathbf{i}}^{\dagger} f_{\mathbf{i}\sigma}$, with
$f_{\mathbf{i}\sigma}$ and $b_{\mathbf{i}}$ being the operators
for a spinon (a neutral spin-$\frac{1}{2}$ fermion) and a holon (a
spinless charged boson). Due to the holon Bose condensation at low
temperatures, the holon operator $b_{\mathbf{i}}$ is treated as a
$c$-number  and the quasiparticles are determined by the spinon
degree of freedom only. Within the mean-field approximation, the
Bogoliubov-de Gennes (BdG) equations can be derived
as~\cite{JXZhu00}:
\begin{equation}
\label{EQ:BdG} \sum_{\bf j} \left( \begin{array}{cc}
H_{\bf ij} & \Delta_{\bf ij} \\
\Delta_{\bf ij}^{\dagger} & -H_{\bf ij}
\end{array} \right) \left( \begin{array}{c}
u_{\bf j}^{n} \\ v_{\bf j}^{n}
\end{array} \right)
= E_{n} \left( \begin{array}{c} u_{\bf i}^{n} \\ v_{\bf i}^{n}
\end{array} \right)\;,
\end{equation}
with
$H_{\mathbf{ij}} = -[t_{\mathbf{ij}}b_{\mathbf{i}}b_{\mathbf{j}}^{*}
+\frac{J_{\mathbf{ij}}}{2}\chi_{\mathbf{ij}}] +
(\epsilon_{\mathbf{i}}-\mu)\delta_{\mathbf{ij}}$   
Here $u_{\bf i}^{n}$ and $v_{\bf i}^{n}$ are the Bogoliubov
amplitudes corresponding to the eigenvalue $E_{n}$. The
resonant-valence-bond (RVB) pairing order parameters (OP)
$\Delta_{\bf ij}$, the bond OP $\chi_{\bf ij}$, and the spatially
dependent hole density $\delta_{\mathbf{i}}$ are determined
self-consistently:
$\Delta_{\mathbf{ij}}= \frac{J_{\mathbf{ij}}}{2}\sum_{n}
[u_{\mathbf{i}}^{n}v_{\mathbf{j}}^{n*}
+u_{\mathbf{j}}^{n}v_{\mathbf{i}}^{n*}] \tanh
\biggl{(}\frac{E_{n}}{2k_{B}T}\biggr{)}$   
$\chi_{\bf ij} = \sum_{n}\{ u_{\bf i}^{n*}u_{\bf j}^{n}f(E_{n})
+v_{\bf i}^{n}v_{\bf j}^{n*}[1-f(E_n)]\}$  
and
$\delta_{\mathbf{i}} =  1-2\sum_{\mathbf{i},n}\{\vert
u_{\mathbf{i}}^{n}\vert^{2} f(E_n)+\vert v_{\mathbf{i}}^{n}
\vert^{2} [1-f(E_n)]\}$    
where $k_{B}$ is the Boltzmann constant; $f(E) =
[\exp(E/k_{B}T)+1]^{-1}$ is the Fermi distribution function. The
holon $c$-number is determined through
$b_{\mathbf{i}}=\sqrt{\delta_{\mathbf{i}}}$. Here without the loss
of generality, we have assumed $b_{\mathbf{i}}$ to be real.
With exact diagonalization, we solve the BdG equations fully
self-consistently. Once a converged solution is reached, the LDOS at
zero temperature is then evaluated according to:
$\rho_{\mathbf{i}}(E) = 2\sum_{n}[\vert u_{\mathbf{i}}^{n}\vert^{2}
\delta(E_{n}-E) +\vert v_{\mathbf{i}}^{n}\vert^{2}\delta(E_{n}+E)]$
where a factor $2$ arises from the spin sum. The LDOS
$\rho_{\mathbf{i}}(E)$ is proportional to the local differential
tunneling conductance which can be measured in an STM
experiment~\cite{MTinkham75}.

\begin{figure}[th]
\centerline{\psfig{file=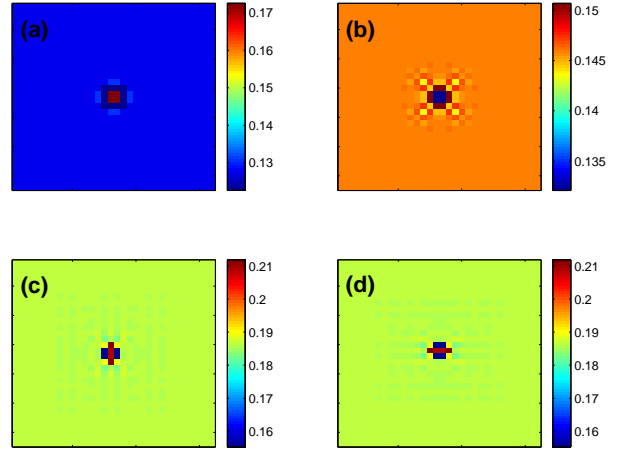,width=8cm}} \caption{ (a) $d$-wave
OP; (b) hole density distribution; (c)-(d) $x$- and $y$ -oriented
RVB bond OP. They are obtained from a solution to an untied single
dopant.
}
\label{FIG:fig1}
\end{figure}

\begin{figure}[th]
\centerline{\psfig{file=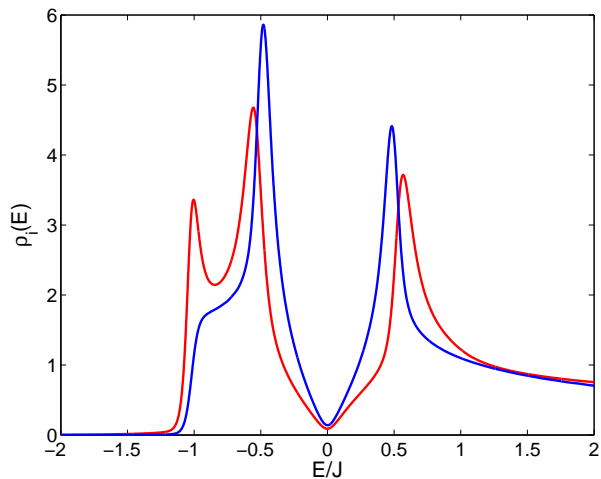,width=8cm}} \caption{LDOS spectrum
at the ``on-dopant'' site with a gap $\Delta_1 \sim 0.6$ (red line)
 and in a location far away from the dopant with a gap
$\Delta_0 \sim 0.5$ (blue line).
}
\label{FIG:fig2}
\end{figure}

\begin{figure}[th]
\centerline{\psfig{file=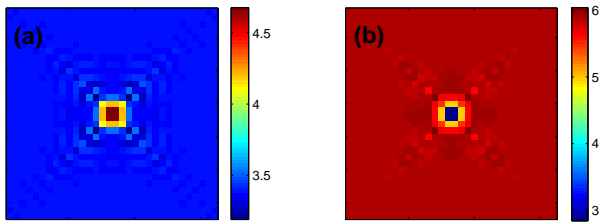,width=8cm}}
\caption{LDOS maps
$\rho_{\mathbf{i}}(E)$ in real space for energy at $E=\Delta_{1}$
(a) and $E=\Delta_0$ (b).
}
\label{FIG:fig3}
\end{figure}

In the numerical calculation, we construct a superlattice with the
square lattice $N_x\times N_y$ as a unit supercell. As detailed in
Ref.~\cite{JXZhu99}, this method can provide the required energy
resolution. Throughout the work, we take the size of the unit
supercell $N_a=32\times 32$, the number of supercells $N_c=6\times
6$, the temperature $T=0$, the Hubbard interaction $U=10$. As we
have already mentioned before, the hopping integral $\tilde{t}_{ij}$
should be modified by the presence of out-of-plane oxygen dopants.
Without a detailed first-principle calculation, the extent of this
modification is unknown. Therefore, we parameterize the hopping
integral in the form~\cite{JXZhu03}:
$t_{\mathbf{ij}}=t_{\mathbf{ij}}^{0}+ \delta t_{\mathbf{ij}}$, where
$t_{\mathbf{ij}}^{0}$ is the bare hopping integral. We choose
$t_{\mathbf{ij}}^{0}$ to be $t=1$ for the nearest-neighboring
hopping and $t^{\prime}=-0.3$ for the next nearest-neighboring
hopping. For the resultant superexchange interactions, only those
for the nearest-neighbor pairs are considered, with a new energy
scale $J=4t^{2}/U=0.4$. By assuming that dopant atoms are
interstitial, as suggested by the STM observations
\cite{ACFang05,KMcElroy05}, we restrict for simplicity the change in
the hopping integral to the four bonds forming a plaquette in the
Cu-O plane, at the location sharply below each oxygen dopant and
take $\delta t_{\mathbf{ij}}=0.2$. No site-diagonal disorder are
introduced, i.e., $\epsilon_{\mathbf{i}}=0$. The chemical potential
is tuned such that the averaged hole density from the
self-consistency solution is about 0.14. The obtained spatial
variation of the $d$-wave is defined as $\Delta_{d}({\bf
i})=\frac{1}{4}[\Delta_{\hat{x}}({\bf i}) +\Delta_{-\hat{x}}({\bf
i}) -\Delta_{\hat{y}}({\bf i}) -\Delta_{-\hat{y}}({\bf i})]$.
Throughout the following discussion, the energy is measured in units
of $J$.

{\em a single dopant atom.}  For the purpose of elucidating the
proposed picture, this calculation is performed without anchoring
the dopant concentration to the averaged hole density. We find that
both the local pair potential (i.e., $d$-wave OP, see
Fig.~\ref{FIG:fig1}(a)) is enhanced on the sites belonging to the
plaquette directly below the dopant atom (for brevity, we will call
these sites as ``on-dopant'' sites). This modulation occurs locally
and the ``healing length'' to recover the bulk value is about the
bulk superconducting coherence length $\xi \sim 5a$. Associated with
the local hopping matrix element modulation, the hole density also
changes. Directly on the off-diagonal impurity, the hole density is
depressed and exihibits the Friedel-like oscillations away from the
dopant site. It means the corresponding Bose-Einstein condensed
holon field is depressed at the dopant site. In addition,
Fig.~\ref{FIG:fig1} (c-d) show the corresponding changes in the $x$-
and $y$- oriented RVB bond OP, which are also recovering a bulk mean
field solution at a few lattice spacings.


The LDOS is shown in Fig.~\ref{FIG:fig2}. The LDOS calculated at the
``on-dopant'' sites clearly shows an enhanced gap with a $V$-shaped
bottom. Interestingly, it is found that the lost spectral weight at
low energy is transferred to the band edge rather than to enhance
the height of the coherent peaks at the gap edge. The ``on-dopant''
gap $\Delta_1$ is larger than the bulk value of the gap $\Delta_0$.
Gap enhancement is about $20\%$ for our choice of parameters.
Enhancement clearly depends on the extent to which we increase a
local hopping matrix element. However, the present result has been
sufficient to convey that the gap modulation as measured by the STM
experiment is proportional to the local pair potential modulation
due to the change in local hopping integral of electrons. It is
qualitatively different from the case of a single site-diagonal
unitary impurity, where the impurity-site LDOS does not show the
shift of the gap edge though the $d$-wave OP has been strongly
suppressed~\cite{AVBalatsky04}.  The contrast between locally large
gap $\Delta_1$ and the sample bulk gap $\Delta_0$ is shown in
Fig.~\ref{FIG:fig3}. For the voltage bias at $E = \Delta_1$, the
only bright spot is on the location corresponding to the dopant
atom. For the bias at $E = \Delta_0$, the bulk of the sample is
bright and the only dark region is on the dopant site.

\begin{figure}[th]
\centerline{\psfig{file=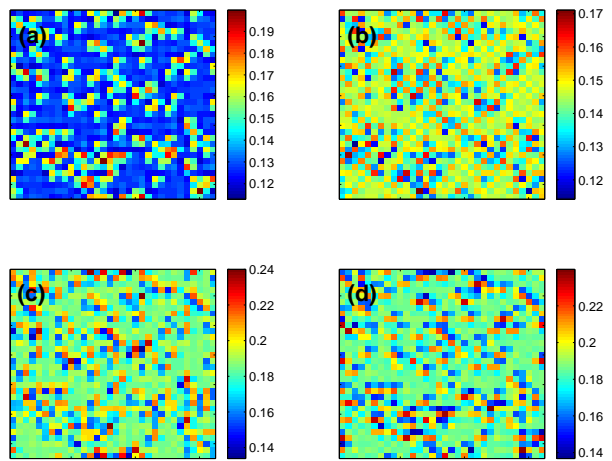,width=8cm}} \caption{Same as
Fig.~\ref{FIG:fig1} but for a concentration of $14\%$ dopants. Here
we have assumed that each dopant atom produces one hole. The
distribution of dopant atoms is random.
}
\label{FIG:fig4}
\end{figure}

\begin{figure}[th]
\centerline{\psfig{file=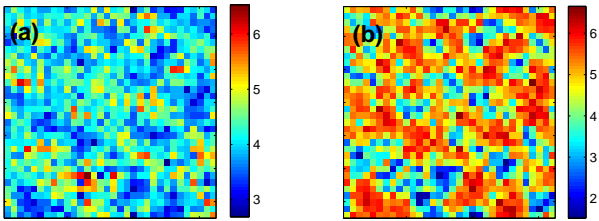,width=8cm}} \caption{LDOS
$\rho_{\mathbf{i}}(E)$ map at $E = \Delta_1$ (a) and $E = \Delta_0$
(b) for a concentration of $14\%$ dopant atoms. } \label{FIG:fig5}
\end{figure}

{\em many dopant atoms.} To strengthen our proposal of
dopant-induced pairing interaction modulation, we turn to the case
of multiple dopants, which is more relevant to the experimental
situation. Here we assume that each oxygen dopant contribute one
hole to the Cu-O plane. Therefore, we tie the concentration of
dopant to the average hole density in the Cu-O plane, i.e., about
14\%. As shown Fig.~\ref{FIG:fig4}, the large pair potential regions
are locally registered to the dopant positions, which are randomly
distributed. The corresponding hole density map shows the variations
that are anti-correlated with the large gap regions. Local $x$- and
$y$- oriented RVB OP also are registered to the dopant locations.
Figure~\ref{FIG:fig5} shows the LDOS map at the voltage bias at
$E=\Delta_1$ and $\Delta_0$, respectively. It demonstrates again
that the intensity map at $E=\Delta_1$ exhibits a positive
correlation with the locally large pair potential regions (the mixed
color areas in \ref{FIG:fig4}(a)) while that at $E=\Delta_0$
reflects the small pair potential regions (the blue areas in
\ref{FIG:fig4}(a)).

In conclusion, we have argued that the local gap inhomogeneity can
be modeled by the modulation in the local superexchange interaction,
induced by the dopant atoms. In the proposed strongly correlated
electronic model, the local pairing interaction is controled by the
superexchange interaction. Therefore, in the presence of the
interstitial dopant atoms, the strongest effect of the doping would
be an off-diagonal disorder that modulates superexchange on the
neighboring bonds. Changes in the local superexchange lead to the
local changes in the pairing strength and, in turn, in the local
magnitude of the $d$-wave gap. An negative correlation of the hole
density variation with the dopant location is also found. The
self-consistent calculations also shows that the regions of large
gap are registered to the dopant atoms. Large gap regions exhibit
smaller and broader coherence peaks. These results are qualitatively
consistent with recent STM observations
by~\cite{McElroy05,ACFang05}, such as positive correlation between
large gap regions and oxygen dopant positions. For a reasonable
choice of parameters, we find the gap variations  by about $20\%$.
Although we use a specific RVB-like model, we believe the effect of
local gap increase to be a general property of any model, where the
 pairing interaction is set by local superexchange.
The following remarks are in order: (i) Our slave-boson mean-field
calculation shows an increase of the average gap with a decreased
doping, where the dopant concentration becomes more dilute. The
result is reasonably consistent with the experimental
data~\cite{KMcElroy05}. However, at this moment, we are unable to
clarify whether the large gap in the low doping regime is dominated
by the proximity to competing orderings; (ii) Although the
unscreened electrostatic potentials generated by the oxygen dopants
do not tune the local pairing interaction in the proposed model,
they might still play a role in the local electronic structure;
(iii) To fully address various aspects of nanoscale electronic
inhomogeneity in high-$T_c$ cuprates, a minimal effective model
should take into account both the site-diagonal and off-diagonal
disorder generated by the oxygen dopants, the resultant modulation
in the superexchange interaction associated with the off-diagonal
disorder, and possibly the residual Hubbard interaction.

{\bf Acknowledgments:} We thank A. V. Balatsky, J. C. Davis, and  K.
McElroy for stimulating discussions and correspondence. We are
especially grateful to A. V. Balatsky for help with the manuscript.
This work was supported by the US DOE.

\end{document}